\documentclass[fleqn,10p]{wlscirep}
\usepackage[utf8]{inputenc}
\usepackage[T1]{fontenc}

\usepackage{graphicx}  
\usepackage{dcolumn}   
\usepackage{bm}        
\usepackage{amssymb,amsfonts,amsmath, color}
\usepackage{float}

\title{Screening by changes in stereotypical behavior during cell motility}

\author[1,2]{Luke Tweedy}
\author[3]{Patrick Witzel}
\author[3, 4]{Doris Heinrich}
\author[2]{Robert  H. Insall}
\author[1, *]{Robert G. Endres}

\affil[1]{Department of Life Sciences and Centre for Integrative Systems Biology and Bioinformatics, Imperial College, London, United Kingdom}
\affil[2]{CRUK Beatson Institute, Glasgow G61 1BD, Scotland, UK}
\affil[3]{Fraunhofer Institute for Silicate Research ISC, Neunerplatz 2, 97082, W{\"u}rzburg, Germany}
\affil[4]{Leiden Institute of Physics, LION, Leiden University}
\affil[*]{Corresponding Author}

\begin{abstract} 
Stereotyped behaviors are series of postures that show very little variability between repeats. They have been used to classify the dynamics of individuals, groups and species without reference to the lower-level mechanisms that drive them. Stereotypes are easily identified in animals due to strong constraints on the number, shape, and relative positions of anatomical features, such as limbs, that may be used as landmarks for posture identification. In contrast, the identification of stereotypes in single cells poses a significant challenge as the cell lacks these landmark features, and finding constraints on cell shape is a non-trivial task. Here, we use the maximum caliber variational method to build a minimal model of cell behavior during migration. Without reference to biochemical details, we are able to make behavioral predictions over timescales of minutes using only changes in cell shape over timescales of seconds. We use drug treatment and genetics to demonstrate that maximum caliber descriptors can discriminate between healthy and aberrant migration, thereby showing potential applications for maximum caliber methods in automated disease screening, for example in the identification of behaviors associated with cancer metastasis.
\end{abstract}
\begin{document}
\flushbottom
\maketitle
\thispagestyle{empty}

\section*{Introduction}
Moving in the right way at the right time can be a matter of life and death. Whether avoiding a predator or searching for food, choosing the correct movements in response to specific stimuli is a crucial part of how an organism interacts with its environment. The repetitive, highly coordinated movements that make up behavioral stereotypes have been shown to be entwined with survival strategies in a number of species, for example the incredible correlation in posture between prey capture events in raptors \cite{csermley1989} and the escape response of \emph{C. elegans} when exposed to extreme heat \cite{Ryu2008}. Understanding these stereotypes is vital to creating a full picture of a species' interactions with its environment. If stereotypes represent evolved, selection-driven behavior in animals, might the same not be true for single-celled organisms?

This point of view may be particularly useful in understanding chemotaxis, the guided movement of a cell in response to a chemical gradient. During chemotaxis, eukaryotic cells change their shape through the repeated splitting and extension of actin-rich structures called pseudods \cite{AndrewAndInsall2007, Neilson2011, Haastert2010}. Though this behavior is well known, the study of chemotaxis has traditionally focused on the signaling events that regulate cytoskeletal remodeling. Even where pseudopods are acknowledged to be relevant, the focus is on the biochemical mechanisms that generate and regulate them \cite{Otsuji2010, Beta2013, Davidson2017}. These mechanisms are, however, staggeringly complex \cite{Devreotes2010} and the way chemotaxis emerges from these lower-level processes remains largely unknown. Rather than delving deeper into the network of biochemical interactions, we can instead learn from the shape changes and movements that this intricate machine has evolved to produce. Such an approach, also known as morphological profiling, shows great promise in biomedicine \cite{Marklein2017}.

Here, we explore this question using \emph{Dictyostelium discoideum}, a model for chemotaxis noted for its reproducible directional migration towards cyclic adenosine monophosphate (cAMP) \cite{vanPost2007, Tweedy2016}, which it senses using typical G-protein coupled receptors. To capture cell shape (or posture) at any given point in time, we employ Fourier shape descriptors, a rotationally invariant method of quantifying shapes used previously to show that cell shape and environment are intrinsically linked \cite{Tweedy2013} (Fig. 1A). These shape data are naturally of high dimensionality, making further analysis difficult. We reduce their dimensionality using principal component analysis (PCA), a method used previously to obtain the key directions of variability from the shapes of cells \cite{Tweedy2013, Keren2008, Bakal2013} and animals (Fig. 1B) \cite{Ryu2008, Broekmans2016, Gyenes2016}. Our final challenge (and the focus of this paper) is to quantify behavior, which we define as the movement between shapes. There are many potential ways to do so \cite{Valetta2017, Gomez2016}, however we have adapted the variational maximum caliber (MaxCal) approach \cite{Phillips2010, Dill2013} to this end. These methods have several advantages over conventional alternatives: Firstly, Fourier descriptors capture all available information on shape, and the subsequent PCA provides a natural quantitative means of discarding its less informative elements. Easier methods, such as measuring aspect ratio or eccentricity, require us to assume that they provide a useful description \emph{a priori}, and cannot inform us how much (or what) we have discarded for the sake of simplicity. Secondly, our chosen methods are blind to the researcher's preconceptions, as well as to previous descriptions of shape and behavior. Any behavioral modes identified have emerged from the data without our deciding on (and imposing) them, as we might if using supervised machine learning or fitting parameters to a preconceived biochemical model. Finally, the minimal model we construct using maximum caliber makes no reference to any underlying biochemistry and therefore cannot make potentially incorrect assumptions about it. We demonstrate the usefulness of these methods by showing that they successfully discriminate between the behavior of drug-treated or genetically altered cells and their parental strains.

\section*{Results}

\subsection*{Maximum caliber approach to behavioral classification}
	
Cells continuously change shape as they migrate, creating trajectories in the space of shapes that are specific to their situation. For example, we have previously shown that cells follow different shape trajectories in environments with low and high chemoattractant signal-to-noise ratios \cite{Tweedy2013}, here defined as the local gradient squared over the background concentration (Fig. 1C).  In this example, it is important to note that the distributions of cell shape for each condition overlap significantly. This means that it is not always possible to accurately determine the cell's condition from a static snapshot of its shape. In contrast, the dynamics of shape change in each condition are clearly distinct. Our aim here is to quantify the details of these shape changes, making a small set of values that can act as a signature for a given mode of behavior. We can then use such signatures to quantitatively compare, or to discriminate between, various conditions or genotypes. To this end, we employ the MaxCal method (Fig. 2A). 

	MaxCal was originally proposed by E. T. Jaynes \cite{Jaynes1980} to treat dynamical problems in physics, and is much like his better-known maximum entropy (MaxEnt) method used in equilibrium physics. The motivation is the same for both; we wish to find the probability of a given structure for a system in a way that neither assumes something we do not know, nor contradicts something we do know, i.e. an observation of the system's behavior. In the case of MaxEnt, this is achieved by finding the maximum Shannon entropy with respect to the probabilities of the set of possible configurations the system can take \cite{Bialek2006}. MaxCal uses the probabilities of the possible trajectories a dynamical system can follow instead. In this case, the entropy is replaced by the caliber \(C\), \cite{Phillips2010}, so called because the flow rate in a pipe relates to its caliber, or internal diameter. In essence, the method extracts the degree to which different rates or events within the system are coupled, or co-occur beyond random chance. This method has previously been used to sucessfully predict the dynamics of neuronal spiking, the flocking behavior of birds and gene regulatory networks \cite{Cavagna2014,Vasquez2012, Firman2017}.

Generally, the caliber takes the form

\begin{equation}
C(\{ p_{j} \}) =  -\sum_{j} p_{j}\ln\left(p_{j}\right) + \mu \sum_{j} p_{j} + \sum_{n} \lambda_{n} \sum_{j} S_{n,j} p_{j} ,
\label{eqn:Cal}
\end{equation} 
where \(p_{j}\) is the (potentially time-dependent) probability of the \(j\)th trajectory. The first term on the right-hand side of Eq. \eqref{eqn:Cal} represents a Shannon-entropy like quantity and the second ensures that the \(p_{j}\) are normalized. The third constrains the average values of some properties \( S_{n, j} \) of the trajectories \( j \) to the values of some macroscopically observed quantities \( \langle S_{n} \rangle \), making sure we do not contradict any known information.

By maximizing the caliber, the probabilities of the trajectories
\begin{equation}
p_{j}  = Q^{-1} \exp \left(\sum_{n} \lambda_{n} S_{n,j} \right)
\label{eqn:MCprobs}
\end{equation}
are found, where \( Q = \sum_{j} \exp(\sum_{n} \lambda_{n} S_{n,j}) \) is the dynamical partition function and \( \{\lambda_{n}\} \) are a set of Lagrange multipliers. This Boltzmann-type distribution fulfils detailed balance, even for non-equilibrium problems. Practically, the problem is to find these Lagrange multipliers (and hence, the partition function). To this end, we exploit their relations to the externally observed average values of some quantities 
\begin{equation}
 \langle S_{n} \rangle = \frac{\partial \ln Q}{\partial \lambda_{n}},
\label{eqn:Observe}
\end{equation}
where the values of the \( \langle S_{n} \rangle \) are determined from experiment. This training process is equivalent to maximum-likelihood fitting to observed data.

As our interest is in cell shape and motility, we derive our values for the \( \langle S_{n} \rangle \) from the shape dynamics of migrating \emph{Dictyostelium} cells. In order to effectively parameterise our model, we must constrain the continuum of possible shape changes to a much smaller set of discrete unit changes in our principal components (PCs). We therefore build our model from discretized values of the shape measures PC 1 and 2, assigning them to the variables \(N_{1}\) and \(N_{2}\), respectively. Their values are analagous to particle numbers in a chemical reaction. The switching between continuous and discrete variables is possible as \(\frac{\sigma_{x}}{\langle N_{x}\rangle}\approx 0.035\) is small with \(x=1,2\) for \(PC_{x}\) and \(\sigma_{x}\) the standard deviation. We reduce the size of the time-step \(\delta t\) until we no longer observe changes greater than 1 in a single \(\delta t\) (similar to deviations of master equations). As PC 2 accounts for less overall variation than PC 1, (see Fig. 1B), we naturally reach this minimal change for a much larger value of \(\delta_t\), which is undesirable because by the time \(\delta_t\) is small enough for PC 1, changes in PC 2 are almost never observed, making correlations between the two PCs difficult to detect. We therefore scale all changes in PC 2 by a factor of \(\sigma_1/\sigma_2\) in order that unit changes are observed in both PCs for a similar value of \(\delta_t\). Practically, our training data yielded a \(\delta_t\) of 0.1875s (as each 3s frame in the video data was divided into 16 sections, in which the intermediate PC values were linearly interpolated).

The advantage of limiting the possible macroscopic shape changes in \(\delta t\) to the following: an increase, a decrease, or no change in each PC. As changes in each PC can be considered independently, this gives us a total of 3x3 = 9 cases (that is, no change form the current position, or a move to any of the 8 neighbouring spaces, see Fig. 2A inset). These macroscopic cases are taken to be the observable effects of an underlying microscopic structure. From our analogy of a chemical reaction, we treat increases to be particle creation and denote the microscopic variable for an increase in trajectory \(j\) as \( S^{x}_{+,j} \), where \(x\in\{1,2\}\) correspond to PC 1 and 2, respectively. For small \(\delta t\) this variable is binary, taking the value \( 1 \) when \(N_{x}\) increases over a single time-step and taking the value \( 0 \) otherwise. Decreases will be treated as particle decay, with \(N_{x}\) separate variables \( \{S^{x,i}_{-,j}\}\) are used to denote decays for the \(i\)th particle, with \(1 \le i \le N_{x}\). These \( \{S^{x,i}_{-,j}\}\) are equal to \( 1 \) if the \(i\)th particle decays in \(\delta t\) and equal to \( 0 \) otherwise. Hence, in each \(\delta t\) there are \(N_{x}+2\) possible microtrajectories for each component; an increase, no change, or the removal of any particle \(N_{x}\) (Fig. 2B). We choose such a first-order decay over a zeroth-order decay in order to introduce a virtual force, bringing the system back toward the mean (see Fig. S2). As the two components may change independently, there are \((N_{1}+2)(N_{2}+2)\) possible microtrajectories in a single \(\delta t\) over PC 1 and 2. Applying Eq. 3, we constrain the probabilities of these microtrajectories such that they agree with the macroscopically observed rates \(\langle S^{x}_{\alpha} \rangle\),  with \(\alpha\in\{+,-\}\) an increase or decrease in component \(x\), respectively. 

We then expand the model to include a following time-step, allowing us to capture short-term correlations between events. This increases the number of possible trajectories substantially. The number of microtrajectories in a given time-step depends on \(N_{x}\) at time \(t+\delta t\), and this quantity is different dependent on the pathway taken in the first time-step, so we must include this history dependence. For example, a reduction in component \(x\) can happen in \(N_{x}\) ways, and will cause \(N_{x}\) to go down by one. This change is followed by \((N_{x}-1)+2\) possible microtrajectories in the following time-step. Multiplying the quantities for each time-step gives us \(N_{x}\big(N_{x}+1\big)\) microtrajectories in which there is a decrease in the first time-step. Accounting for the effect of the changing values of \(N_{1}\) and \(N_{2}\) over interval \({t,t+\delta t}\) in each microtrajectory on the interval starting at time \(t+\delta t\), the number of microtrajectories over \(2\delta t\) is \((N_{1}^{2}+3N_{1}+5)(N_{2}^{2}+3N_{2}+5)\). Each observable has a corresponding value in trajectory \(j\) of \(S^{xy}_{\alpha \beta, j}\), which is 1 if the correlation is observed and 0 otherwise.  We can reduce this to 10 time-correlated observables by assuming symmetry under order-reversal, \emph{i.e.} \(S^{xy}_{\alpha \beta, j} \equiv S^{yx}_{\beta \alpha, j}\) (Fig. 2C). This assumption is justified: if we consider a negatively correlated movement between PC1 and PC2, we may see transitions in the order \(1+, 2-, 1+\). Here the two couplets \(1+,2-\) and \(2-,1+\) both represent the same phenomenon (see Fig S3).
This leads to an additional 16 observables \(\langle S_{xy}^{\alpha \beta} \rangle\), where \(x,y\in\{1,2\}\) are shape PCs and \(\alpha, \beta \in \{-,+\}\) denote a change in the component displayed above. We constrain our analysis to the first two shape components only, as further components account for relatively little residual variance in shape, whilst increasing computational complexity geometrically.

As an example, we show the partition function in a single shape component, in which there are 5 observables,

 \(\{\langle S^{+} \rangle,\langle S^{++} \rangle, \langle S^{+-} \rangle, \langle S^{-} \rangle, \langle S^{--} \rangle\}\): \begin{align}
Q_{N} &=  \gamma^{+}\big[\gamma^{+}\gamma^{++}+1+\big(N+1\big)\gamma^{-}\gamma^{+-}\big] \nonumber\\
         &+N\gamma^{-}\big[\gamma^{+}\gamma^{+-}+1+\big(N-1\big)\gamma^{-}\gamma^{--}\big] \nonumber \\ 
	 &+ \gamma^{+}+1+N\gamma^{-},
\label{eq:exampleQ}
 \end{align} where \(\gamma^{\alpha}=e^{\lambda^{\alpha}}\) corresponds to a rate (when divided by \(\delta t\)), with  \(\lambda^{\alpha}\) the Lagrange multiplier associated with observable \( \langle S^{\alpha} \rangle \). The first line in Eq. \eqref{eq:exampleQ} shows all possible transitions that begin with an increase over the first time-step, and so the whole line shares the factor \(\gamma^{+}\), the rate of increase. A subsequent increase contributes a further \(\gamma^{+}\), as well as a coupling term \(\gamma^{++}\) which allows us to capture the likelihood of adjacent transitions beyond the naive probability \(\gamma^{+} \gamma^{+}\). A subsequent decrease can happen in \(N+1\) ways, each linked to the rate of decrease \(\gamma^{-}\). The term \(\gamma^{+-}\) is a coupling constant, controling the likelihood of an adjacent increase and decrease beyond the naive probability \(\gamma^{+} \gamma^{-}\). Finally, the +1 allows for the possibility of no transition occurring in the subsequent time-step. The second and third lines correspond to a decrease in the first time-step, and no transition occurring in the first time-step, respectively.

The Lagrange multipliers corresponding to observables are found using Eq. \eqref{eqn:Observe}, which yields a set of equations to be solved simultaneously (see supplementary material for details). In the case of a single component, these equations are

 \begin{gather}
\refstepcounter{equation}
\begin{align}
\tag{\theequation a}
\langle S^{+} \rangle &= \gamma^{+}\bigg[2\gamma^{+}\gamma^{++} + 2 + (2N+1)\gamma^{-}\gamma^{+-}\bigg]\\ \tag*{}
\langle S^{-} \rangle &=  \gamma^{-}\bigg[(2N+1)\gamma^{+}\gamma^{+-}+2N \\\tag{\theequation b} &\;\;\;\;\;\;\;\;\;\;\; + 2N(N-1)\gamma^{-}\gamma^{--}\bigg]
\end{align}\\ \begin{align}\tag{\theequation c}
\langle S^{++} \rangle &=  \gamma^{+}\gamma^{+}\gamma^{++}\\ \tag{\theequation d}
\langle S^{+-} \rangle &=  2N\gamma^{+}\gamma^{-}\gamma^{+-}\\  \tag{\theequation e}
\langle S^{--} \rangle &=  N(N-1)\gamma^{-}\gamma^{-}\gamma^{--}.
\end{align}
 \end{gather}
The equations for the two-component partition function and Lagrange multipliers can be found in the SI.
This method effectively allows us to build a map of the commonality of complex, correlated behaviors relative to basic rates of shape change (as quantified using principal components). For a given Lagrange multiplier governing a particular correlation, a value less than zero indicates a behavior that is less common than expected, and a value greater than zero represents a behavior that is more common.
\subsection*{Stereotypical behavior without biochemical details} 

After training our model on \emph{Dictyostelium} shape trajectories, we confirmed that the method had adequately captured the observed correlations by using them to simulate the shape changes of untreated cells responding to cAMP. In order to illustrate the importance of the correlations, we also ran control simulations trained only on the basic rates of increase and decrease in each PC without these correlations. We compared the activities of the uncorrelated and correlated simulations against the observed data. The uncorrelated model acts entirely proportional to the observed rates (though, interestingly, did not match them; Fig. 2D). In contrast, individual cells from the experimental data show very strong anticorrelation, with increases in one component coupled with decreases in the other. This behavior is clearly replicated by the correlated simulations, in both cases appearing in the plot as a red diagonal from the bottom left to the top right. Furthermore, we see suppression of turning behavior in both PCs, with the most poorly represented activity (relative to chance) being a switch in direction in either PC (for example 1+ followed by 1-). This too is reflected in the correlated simulations. 

The predictive power of MaxCal simulations goes beyond those correlations on which they were directly trained:  We tested the simulations' ability to predict repetition of any given transition. These patterns took the form of \(N\) transitions in \(T\) time steps, e.g. five 1+ transitions in ten time-steps. The MaxCal model predicted frequencies of appearance for these patterns that closely resembled the real data (Fig. 3A, model in red, real data in black).  In contrast, the uncorrelated model predicted patterns at a much lower rate, for example there are runs of 5 consecutive increases in PC 1 in the real data at a rate of around one in 1.35 minutes. The correlated model predicts this pattern rate to be one in every three minutes. The uncorrelated model predicts the same pattern at a rate of one in 6.67 hours. This result indicates that no higher-order correlations are required to recapitulate the data, allowing us to avoid the huge increase in model complexity that their inclusion would entail. 

The greater predictive power of the MaxCal model is reflected by its lower Jensen-Shannon divergence from the observed data for these kinds of pattern (Fig. 3B). The MaxCal model also more closely matches the observed probabilities of generating a given number of transitions in a row, with predictions almost perfect up to 4 transitions in a row (twice the length-scale of the measured correlations), and far stronger predictive power than the uncorrelated model over even longer timescales (Fig. 3C).

\subsection*{A real world application of MaxCal methods to discriminate between genotypes}

We wondered whether the MaxCal methods would accurately discriminate between biologically relevant conditions. To investigate this, we used two comparisons. First, we compared shape data from control AX2 cells against the same cells treated with two drugs targeting cytoskeletal function: the phospholipase A2 inhibitor p-bromophenacyl bromide (BPB) and the phosphoinositide 3-kinase inhibitor LY294002 (LY) (for details, see \cite{Meier2011}). Second, we compared a stable myosin heavy-chain knockout against its parent strain (again AX2) (Fig. 4A). We first looked at the effects of these conditions on the distribution of cell shapes, to see whether their effect could be identified from shape, rather than behavior. The drug treatment caused a substantial change in the distribution within the population (Fig. 4B), but still left a substantial overlap. In contrast, the \( \emph{mhcA}^{-} \) cells showed no substantial difference to their parent in shape distribution (Fig 4C). In both cases, the identification of a condition from a small sample of shape data would not be feasible.

We then compared the behavioral Lagrange multipliers of each condition, found by MaxCal, producing distributions for the estimated values of these by bootstrapping our data (sampling with replacement). The values of \(\gamma_{\:1\: 1}^{+-}\) and \(\gamma_{\:2\: 2}^{+-}\) are lower in the untreated condition than those in drug-treated condition, indicating the persistence of shape change in WT cells (Fig. 4D). The anticorrelation between PCs 1 and 2 through pseudopod splitting is reflected in \(\gamma_{\:1\: 2}^{+-}\) and \(\gamma_{\:1\: 2}^{-+}\), both of which have values greater than 1 in WT cells. In comparison, the drug-treated cells have only a moderate anticorrelation. In the \( \emph{mhcA}^{-} \) strain, the differences in the values of \(\gamma_{\:1\:1}^{+-}\), \(\gamma_{\:1\:2}^{+-}\) and \(\gamma_{\:1\:2}^{-+}\) when compared with their parent show similar changes to those observed in drug treatment (Fig. 4E). In both cases, the differences highlighted by these dynamical measurements are striking. 

We then applied the MaxCal model to the task of classification. We settled upon classification using \emph{k}-nearest-neighbors (kNN). In order to see how the strength of our prediction improved with more data, we classified based on the preferred class of \emph{N} repeats, all known to come from the same data set. We estimated the classification power of our methods by cross validation, dividing the drug-treated data and its control into three sets containing different cells, and dividing the \( \emph{mhcA}^{-} \) and its parent by the day of the experiment. We first performed the classification by shape alone, taking small subsamples of frames from each cell and projecting them into
 their shape PCs, with our distance measure for the kNN being the Euclidean distance in these PCs. With one, two or 3 PCs, we were able to achieve reasonable classification of the drug-treated cells against their parent as data set size increased, with the accuracy of classification leveling off at around 0.85 (with 1 being perfect and 0.5 being no better than random chance, Fig. 5A-C, blue). In contrast, classification of \( \emph{mhcA}^{-} \) cells was little better than random chance, even with relatively many data (Fig. 5A-C, green). This is unsurprising given the similarity of the distributions of these two conditions. We then calculated our MaxCal multipliers for subsamples of each of these groups, bootstrapping 100 estimates from \(20\%\) of each set. We then repeat our kNN classification, instead using for a distance measure the two MaxCal values that best separate our training classes. As the test data come from entirely separated sources (in the case of drug-treated cells coming from different cells, and in the case of the \( \emph{mhcA}^{-} \) being taken on different days), we can be confident that we do not conflate our training and test data. In both the drug-treated case and the \( \emph{mhcA}^{-} \) mutant, the dynamics differ very cleanly between our test and control data. As such, our classification is close to perfect even for only a few samples (Fig. 5D).

As the two Lagrange multipliers that best classified the data both encoded correlations between adjacent time-steps, we guessed that this short-term memory might be key to recapitulating the dynamic properties of cell shape change. A key aspect of the shape dynamics of AX2 cells is the anticorrelation between the first two PCs at the single-cell level (which is definitionally absent at the population level, as PCA produces uncorrelated axes). To see if memory is vital to recapitualting this dynamical aspect of cell shape change, we constructed two versions of the master equation for our MaxCal system (see SI for details). The first is Markovian (that is, at a given time the probabilities of each possible next event only depend on the current state of the system). We ran Gillespie simulations corresponding to this master equation, and compared the correlations of trajectories from these simulations with those from real data. The expected anticorrelation is clearly observed in the data (Fig 5E, black line), but the trajectories of our Markovian Gillespie simulations fail to recapitulate it (Fig. 5E, blue line). 

We then introduced a memory property to the simulations, allowing the probabilities of each possible event to depend on the nature of the previous event (with the very first event taken using the uncorrelated probabilities). The model has nine possible states (with each state containing its own set of event probabilities), corresponding to the nine possible events that might have preceeded it. These are an increase, a decrease, or no change in each PC indepenently (3x3 events).  These non-Marovian simulations recovered the distribution of correlations observed in the data (Fig. 5E, red line). This indicates that such features of cell shape change can only be addressed by methods that acknowledge a dependency on past events. 

\section*{Discussion} 

Eukaryotic chemotaxis emerges from a vast network of interacting molecular species. Here, instead of examining the molecular details of chemotaxis in \emph{Dictyostelium discoideum}, we have inferred properties that capture cell behavior from observations of shape alone. For this purpose, we quantified shape using Fourier shape descriptors, reduced these shape data to a small, tractable dimensionality by principal component analysis, and built a minimal model of behavior using the maximum caliber variational method. Unlike conventional modeling approaches, such as master equations and their simplifications, our method is intrinsically non-Markovian, capturing memory effects and short-term history in the values of the behavioral signature it yields (see SI for further discussion of memory, and a comparison to the master equation). Our approach has the advantage of ease, requiring only the observation of what a cell naturally does \cite{Keren2008}, without tagging or genetic manipulation, as well as of generality, being independent from the specific and poorly understood biochemistry of any one cell type. This is important to understanding chemotaxis, as the biochemistry governing this process can vary greatly: for example, the spermatazoa of \emph{C. elegans} chemotax and migrate with no actin at all \cite{Nelson1982}, but strategies for accurate chemotaxis might be shared among biological systems and cell types.

A number of recent studies have demonstrated the importance of pairwise or short-scale correlations in determining complex behaviors both in space and time. The behavior of large flocks of starlings can be predicted from the interactions of individual birds with their nearest neighbors \cite{Toner1995, Bialek2012}, and the pairwise correlations between neurons can be used to replicate activity at much higher coupling orders, correctly reproducing the coordinated firing patterns of large groups of cells \cite{Bialek2006}. Furthermore, cells in many circumstances use short-range spatial interactions to organise macroscopically \cite{DePalo2017}. Interestingly, these systems appear to exhibit self-organized criticality \cite{Bialek2011, Dante2010}, in which the nature of their short-range interactions leads to periods of quiescence puntuated by sudden changes. This could indicate the coupling strengths inherent to a system (such as the temporal correlations in our shape modes in \emph{Dictyostelium} cells) are crucial for complex behavior. Absence of this behavior could be an indicator of disease as illustrated by both of our aberrant cell types. 

Here, we employ a very simple classifier to demonstrate the usefulness of our MaxCal multipliers as a measurement by which we can classify cell behaviours. We choose MaxCal because it is a minimal, statistical approach to modelling a complex phenomenon, allowing high descriptive power with no assumptions made about the underlying mechanism.
As our understanding of the molecular biology controlling cell shape improves, an interesting alternative would be to use our data in training recurrent neural network (RNN) auto-encoders, a self-supervised method in which the neural network trains a model to accurately represent the input data.  In particular, long short-term memory RNNs have recently been used to accurately identify mitotic events \cite{Phan2018} in cell video data and classes of random migration based on cell tracks \cite{Kimmel2019}. The two approaches are not mutually exclusive; MaxCal can provide a neat, compressed basis in which to identify behavioural states of cells, whilst RNNs could be used to learn time-series rules for transitions between behavioural states.

It is increasingly clear that cell shape is a powerful discriminatory tool \cite{Marklein2017}. For example, diffeomorphic methods of shape analysis have the power to discriminate between healthy and diseased nuclei \cite{Rohde2008}. Shape characteristics can also be used as an indicator of gene expression \cite{Bakal2013}: an automated, shape-based classification of \emph{Drosophila} haemocytes recently showed that shape characteristics correlate with gene expression levels, specifically that depletion of the tumor suppressor PTEN leads to populations with high numbers of rounded and elongated cells. Of particular note is the observation from this study that genes regulate shape transitions as opposed to the shapes themselves, illustrating the importance of tools to quantify behavior as well as shape. This may be an appropriate approach to take if, for example, creating automated assistance for pathologists when classifying melanocytic lesions (a task which has already proved tractable to automated image analyses \cite{Esteva2017}), as classes are few in number, predefined and extensive training data are available. A drawback of the method used by \cite{Bakal2013} is that their classes are decided in advance, and the divisions between them are arbitrary. This means that the method cannot find novel important features of shape by definition, as it can only pick between classes decided upon by a person in advance. 

A stronger alternative would be to take some more general description of shape and behavior (such as the one we detail here), which could be used to give biopsied cells a quantitative signature. Training would then map these data not onto discrete classes, but onto measured outcomes based on the long-term survival of patients. It will be important for any such method to account for the heterogeneity of primary tissue samples as small sub-populations, lost in gross measurements, may be key determinants of patient outcomes. Such an approach would allow a classifier to identify signs of disease and metastatic potential not previously observed or conceived of by the researchers themselves. As machine learning advances, it will be vital to specify the problem without the insertion of our own biases. Then, behavioral quantification will become a powerful tool for medicine.


\section{Methods}
{\bf Cell culture.} The cells used in our experiments are either of the \emph{Dictyostelium discoideum} AX2 strain, or a stable myosin heavy-chain knockout (\( \emph{mhcA}^{-} \)) in an AX2 background. Cells are grown in a medium containing \(10 \mu g/mL\) Geneticin 418 disulfate salt (G418) (Sigma-Aldrich) and \(10 \mu g/mL\) Blasticidine S hydrochloride (Sigma-Aldrich). Cells are concentrated to \( c = 5 \times 10^{6}\) cells\(/mL\) in shaking culture (150 rpm). Five hours prior to the experiment, cells are washed with \(17 mM \) K-Na PBS pH 6.0 (Sigma-Aldrich). Four hours prior to the experiment, cells are pulsed every 6 minutes with 200nM cAMP, and are introduced into the microfluidic chamber at \(c = 2.5 \times 10^{5} \)cells\(/mL\). Measurements are performed with cells starved for 5-7 h. Drug-treated cells were exposed to \(200pM\) p-bromophenacyl bromide and \(50nM\) LY294002. No. cells sampled in AX2 control, drug-treated, AX2 parent, \( \emph{mhcA}^{-} \) are, respectively, 313, 23, 858,198. 

{\bf Microfluidics and imaging.} The microfluidic device is made of a \(\mu\)-slide 3-in-1 microfluidic chamber (Ibidi) as described in (12), with three \(0.4 \times 1.0 mm^{2}\) inflows that converge under an angle of \(\alpha = 32^{\circ}\) to the main channel of dimension \(0.4 \times 3.0 \times 23.7 mm^{3}\). Both side flows are connected to reservoirs, built from two \(50 ml\) syringes (Braun Melsungen AG), separately connected to a customized suction control pressure pump (Nanion). Two micrometer valves (Upchurch Scientific) reduce the flow velocities at the side flows. The central flow is connected to an infusion syringe pump (TSE Systems), which generates a stable flow of \(1 ml/h\). Measurements were performed with an Axiovert 135 TV microscope (Zeiss), with LD Plan-Neofluar objectives \(20x/0.50 N.A.\) and \(40x/0.75 N.A.\) (Zeiss) in combination with a DV2 DualView system (Photometrics). A solution of \(1 \mu M\) Alexa Fluor 568 hydrazide (Invitrogen) was used to characterize the concentration profile of cAMP (Sigma-Aldrich) because of their comparable molecular weight. 

{\bf Image preprocessing.} We extracted a binary mask of each cell from the video data using Canny edge detection, thresholding, and binary closing and filling. The centroid of each mask was extracted to track cell movement. Overlapping masks from multiple cells were discarded in order to avoid unwanted contact effects, such as distortions through contact pressure and cell-cell adhesion. For each binary mask, the coordinates with respect to the centroid of 64 points around the perimeter were encoded in a complex number, with each shape therefore recorded as a 64 dimensional vector of the form \({\bf S} = {\bf x} + i{\bf y}\). These vectors were passed through a fast Fourier transform in order to create Fourier shape descriptors. Principal component analysis was performed on the power spectra (with the power spectrum \(P(f) = |s(f)|^{2}\) for the frequency-domain signal \(s(f)\)) to find the dominant modes of variation. This approach is superior to simple descriptors such as circularity and elongation, as key directions of variability within the high-dimensional shape data cannot be known a-priori. As we have previously reported \cite{Tweedy2013}, 90\% of \emph{Dictyostelium} shape variation can be accounted for using the first three principal components (PCs), corresponding to the degree of cell elongation (PC 1), pseudopod splitting (PC 2) and polarization in the shape (PC 3) (Fig. 1B), with around 85\% of variability accounted for in just two, and 80\% in one. \newline 

\section{Acknowledgements}
We are grateful to B\"{o}rn Meier for sharing his data, and to both Andr\'{e} Brown, Linus Schumacher and Peter Thomason for a critial reading of the manuscript. This work was supported by Cancer Research UK core funding (L.T.), the Deutsche Forschungs-gemeinschaft (DFG fund HE5958/2-1), the Volkswagen Foundation grant I/85100 (D.H), and the BBSRC grant BB/N00065X/1 (R.G.E.) well as the ERC Starting Grant 280492-PPHPI (R.G.E.). 

\section*{Author contributions statement}
LT and RGE designed the study. LT and PW performed the experiments, and LT conducted 
data analysis and modelling. All authors (LT, PW, DH, RHI, RGE) analyzed results and data, 
and wrote the paper. 

\section*{Additional information}
\textbf{Competing interests}
All authors declare that there is no conflict of interest, neither financial nor non-financial.

\begin{figure}[h]
\begin{center}
\includegraphics[width=.48\textwidth]{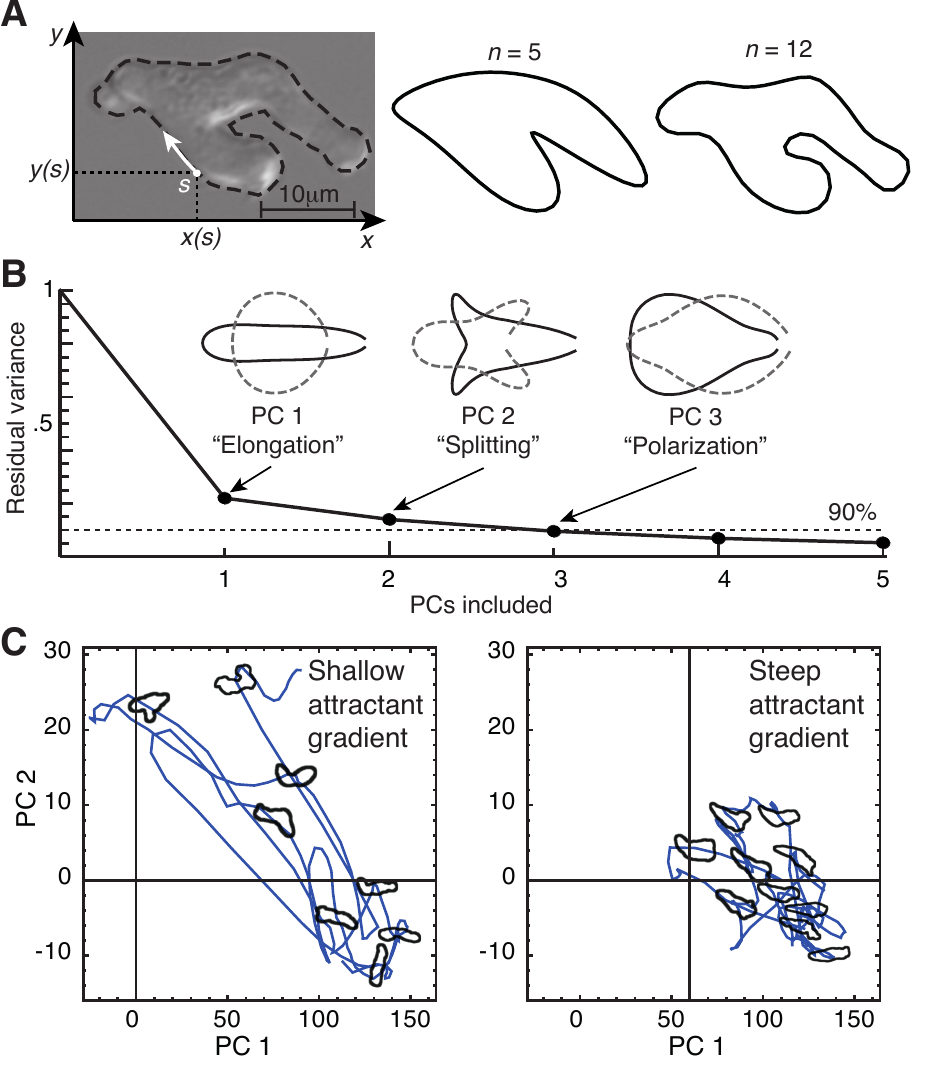}
\caption{{\bf Fourier shape descriptors reveal low-dimensional shape space.} {\bf(A)} (left) The outline of a live cell is converted to a set of coordinates, which are a function of the distance traveled around the perimeter, \(s\). (right) The original outline can be reconstructed with increasing accuracy by increasing the number  \(n\) of included Fourier components. We record 64 components for each shape and use all of them in our analyses. {\bf(B)} Principal component analysis (PCA) performed on Fourier spectra of cell shapes from 900 cells in a wide range of chemical gradients reveals that 90(83)\% of shape variability in \emph{D. discoideum} can be accounted for in the first three (two) principal components, corresponding to elongation, splitting and polarization in the spatial domain. The inset picture above each PC shows its reconstruction in the spatial domain (i.e. after reverse Fourier transformation). Each is added to (solid line) and subtracted from (dashed line) the mean cell shape descriptor. In order to guarantee their invariance when rotated or flipped we use only the power spectrum of the Fourier component, which renders their reconstructions symmetric. We show shapes here that are two standard deviations above (solid lines) and below (dashed lines) the mean shape in each PC. For Fourier contributions to each PC, see Fig. S1. For details on data collection and analysis, see Materials and Methods and \cite{Tweedy2013}. {\bf(C)} Example trajectories in the PC 1 and PC 2 shape space for one low- and one high-signal-to-noise ratio cell. Example cell outlines from the two trajectories are superimposed in their correct positions.} 
\label{fig:shapeIntro}
\end{center}
\end{figure} 

\begin{figure}[h]
\begin{center}
\includegraphics[width=.96\textwidth]{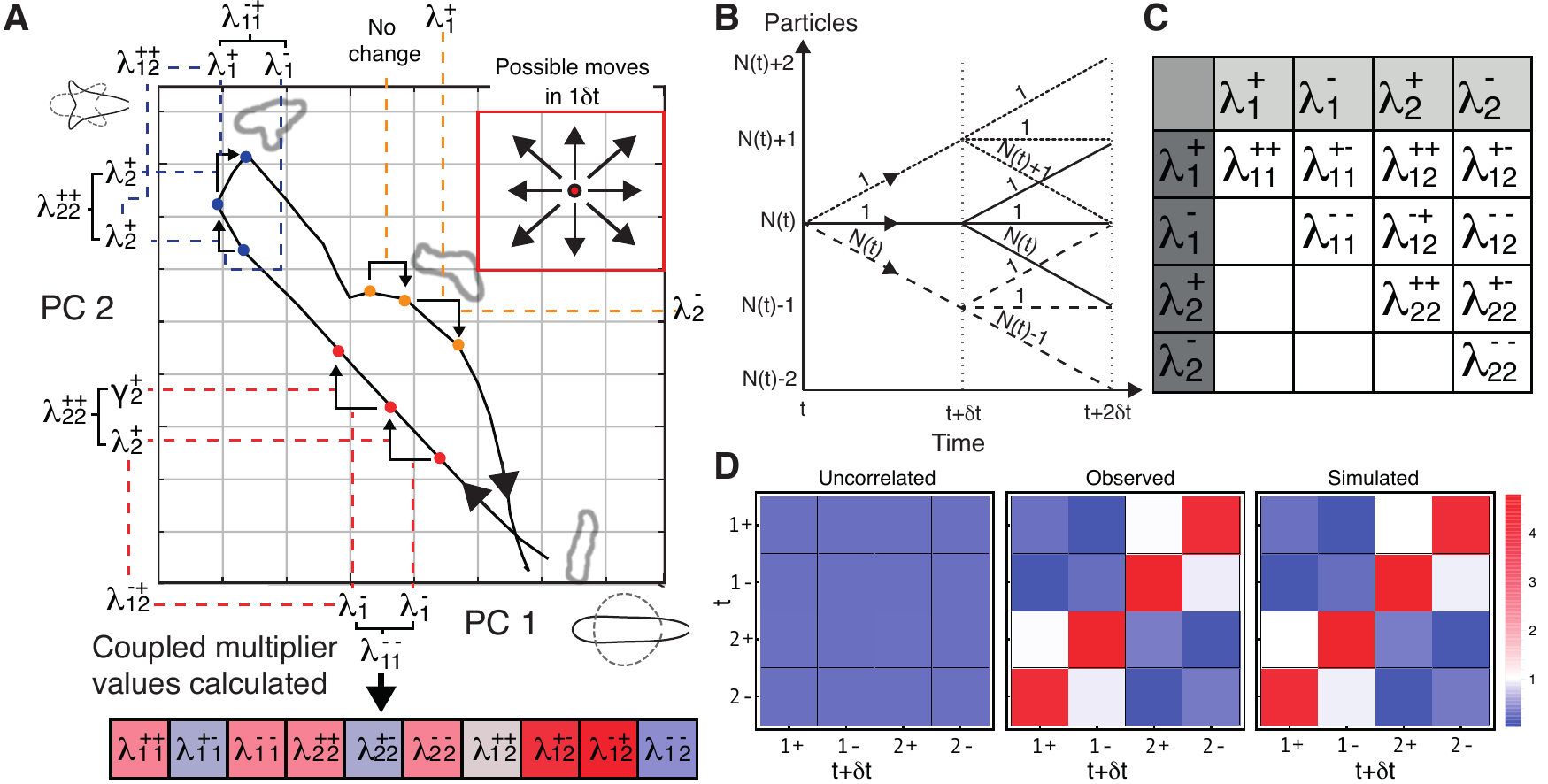}
\caption{{\bf MaxCal trained simulations reproduce local correlations.}  {\bf(A)} The panel shows the trajectory of a cell in shape space over time as it shortens, splits pseudopods, commits to one pseudopod and lengthens again. Our aim is to distil this complex behavioral information into a small, quantifiable signature for this behavor in a manner that will yield similar signatures for similar behaviors. We subdivide the shape space, and register specific small events when cells cross boundaries. The elements of our signature are a series of multipliers that are determined by the rates at which particular events are observed. We introduce multipliers for two types of events: simple events, which encode the average rate of increase or decrease in each principal component of shape, and correlated events, which encode how often we see certain combinations of simple events in quick succession. For every pair of simple events there is a specific multiplier (see C). As multipliers for simple events encode the average rate at which transitions between squares occur in any particular direction, the input data for these events are any two adjacent time points, for example the first two red spots. Here we see that PC 1 has decreased, increasing the average value of the multiplier \(\lambda_{1}^{-}\), and PC 2 has increased, thereby increasing the value of \(\lambda_{2}^{+}\). In contrast, the first two yellow points occupy the same square, and as we are encoding the likelihood of transitions, the values of all simple event multipliers are lowered to represent this inactivity. Multipliers for correlated events involve three adjacent time points, as they encode the likelihood of any one simple event following another particular simple event. For example, the three red dots increase the value of three correlated multipliers. \(\lambda_{\:1\:1}^{--}\) due to a decrease in PC 1 being adjacent to another decrease in PC 1, \(\lambda_{\:2\:2}^{++}\) due to an increase in PC 2 being adjacent to another increase in PC 2, and \(\lambda_{\:1\:2}^{-+}\) due to a decrease in PC 1 being adjacent to an increase in PC 2. The set of yellow dots does not increase the likelihood of any such correlated events as there are no transitions in the first time step. The contributions of the whole trajectory can eventually comprise a quantitative signature in which correlated events appear more or less frequently than would be expected by chance (shown below the main diagram, with blue indicating a higher (and red a lower) likelihood. The calculation of the magnitudes of these contributions is complicated and makes up much of this paper. Note that, though the example trajectory shown here is real, the sub-divisions and time points are instructional only. ({\bf Inset}) In a single timestep (\(1\delta t\)), a cell may be observed to move to any of the eight neighbouring squares, or may move to none of them, resulting in nine possible observed trajectories. {\bf(B)} Diagram of all possible trajectories for a single shape component over time interval \(2\delta t\). The redundancy of each path is indicated by the number above. {\bf(C)} Correlation parameters inferred from data. Lagrange multipliers are found corresponding to increases (\(\lambda_{1}^{+}\)) and decreases (\(\lambda_{1}^{-}\)) in PC 1 and increases (\(\lambda_{2}^{+}\)) and decreases (\(\lambda_{2}^{-}\)) in PC 2. Additional Lagrange multipliers controlling the rates at which these events occur in neighboring time-steps are shown in the table, and rates are assumed to be symmetric (i.e. event A followed by event B is equivalent to event B followed by event A).  {\bf(D)} Co-occurrences of transitions in neighboring time-steps are shown for simulations based only on naive probabilities (without correlations, left), for data (centre) and for simulations that include short-term temporal correlations (right). All are scaled relative to the naive rates of transition observed in the data.}
\label{fig:shapeIntro}
\end{center}
\end{figure} 

\begin{figure}[t]
\begin{center}
\includegraphics[width=.48\textwidth]{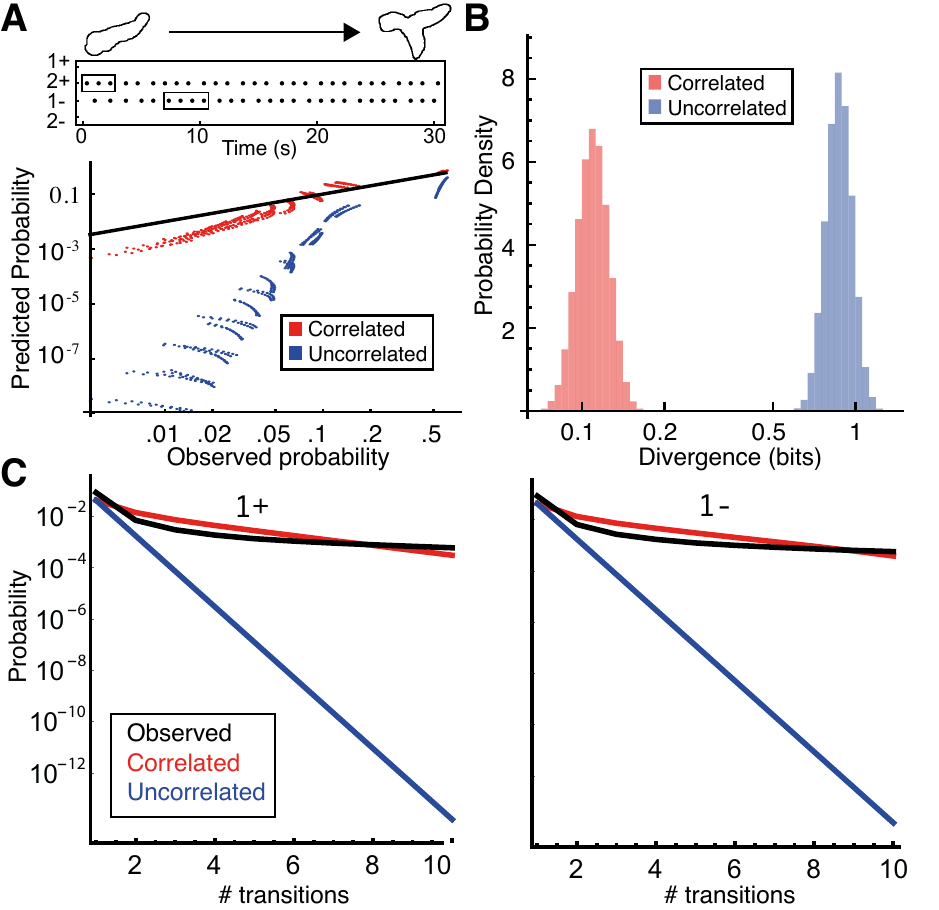}
\caption{{\bf MaxCal model reproduces long-term cell behavior} {\bf(A)} (top) Example track of naive transitions. Possible patterns are highlighted: three 2+ in a row and four 1- in a row are given as examples. This panel is illustrative- though the track shown does correspond to the shape transition in the cartoon, there are spaces in between all transitions here, and actual 3-in-a-row transitions would be even denser. (Bottom) Observed versus predicted probabilities of various patterns of transitions for simulations containing (red) or ignoring (blue) short-term temporal correlations. The black line corresponds to perfect matching of predicted and observed probabilities. Patterns are always for a single transition type (e.g. 1+ only), and are of the form ``N transitions in T time-steps'' for varying N and T, e.g. three 1+ transitions in four time-steps. T runs from 10 to 20 in unit steps, N runs from T-8 to T in unit steps. {\bf (B)} Jensen-Shannon divergence between patterns of transitions observed in the data and both correlated (red) and uncorrelated (blue) simulations. {\bf (C)} Probability of observing repeated transitions of a single type under 3 models- the correlated model (red), the uncorrelated model (blue), and when observed (black). Results shown for 1+ (left) and 1- (right). The x-axis gives the number of repeats in a row of this transition.}
\label{fig:shapeIntro}
\end{center}
\end{figure}

\begin{figure}[h]
\begin{center}
\includegraphics[width=.9\textwidth]{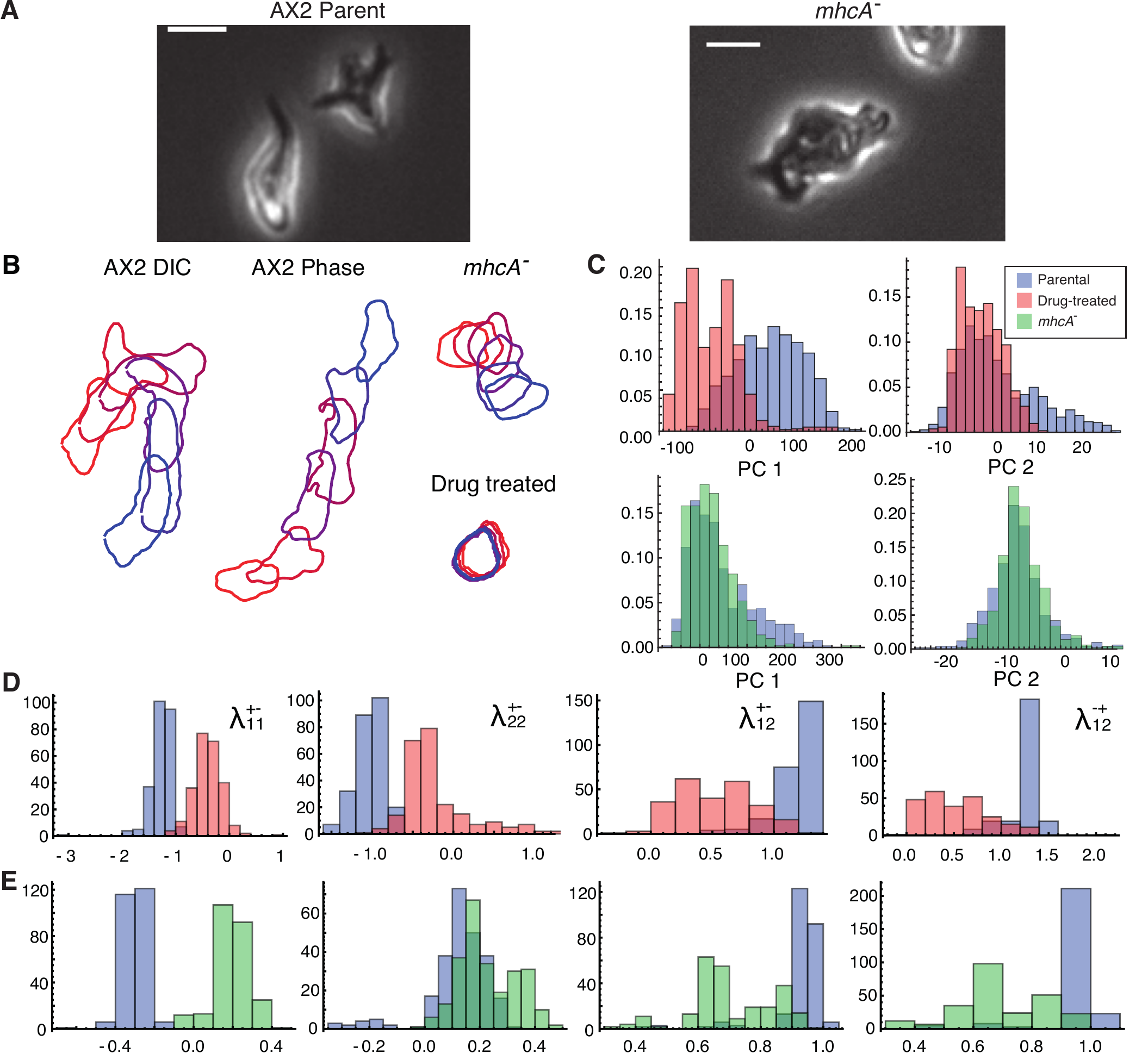}
\caption{{\bf Maximum caliber coupling strengths reflect behavioral differences.}  {\bf(A)} Representative phase contrast images of AX2 (L) cells and \( \emph{mhcA}^{-} \) (R) cells. Scale bar is \(10\mu m\). {\bf(B)} Outlines of cells as they move for all conditions in the study. We imaged drug-treated cells paired with an AX2 parental control in differential interference contrast (DIC) and the \(mhcA^{-}\) control with an AX2 parent in phase contrast microscopy, so we present these controls separately. Here, for clarity, we show outlines at intervals of one minute (or thirty frames). {\bf(C)} Distributions in shape space of the BPB/LY treated cells (red) and the \( \emph{mhcA}^{-} \) strain (green) vs their respective controls (blue). Darker red and green areas show where the compared conditions overlap. {\bf (D)} Lagrange multipliers \(\lambda_x = \text{log}_{10}\:  \gamma_x\) from MaxCal model trained on untreated cells (blue) and p-bromophenacyl bromide (BPB) and LY294002 treated cells (red). Untreated cells have coupling values for \(\lambda_{\:1\:1}^{+-}\) and \(\lambda_{\:2\:2}^{+-}\) much lower than 0, indicating the persistence in the direction of cell shape change (or more accurately, the rarity of reversals). In contrast, treatment with  causes cells to have couplings values for \(\lambda_{\:1\:1}^{+-}\) and \(\lambda_{\:2\:2}^{+-}\) that are not significantly different to zero, indicating a lack of such persistence. Untreated cells also have much higher values than drug-treated for \(\lambda_{\:1\:2}^{+-}\) and \(\lambda_{\:1\:2}^{-+}\), indicating a stronger anticorrelation between the two components. Remaining distributions of rates can be found in Fig. S4. The distributions of all 14 rates differ significantly upon drug treatment \(\alpha=0.001\) (Kolmogorov-Smirnov test). {\bf (E)} The distributions for the same parameters for \( \emph{mhcA}^{-} \) cells (green) against their parental control (blue). The changes in \(\lambda_{\:1\:1}^{+-}\), \(\lambda_{\:1\:2}^{+-}\) and \(\lambda_{\:1\:2}^{-+}\) show similar changes to those observed in drug treatment. These experiments used phase contrast microscopy rather than the DIC used in the drug-treated case, leading to the differences in the extremity of these coupling coefficients. }
\label{fig:MaxCalDrugs}
\end{center}
\end{figure}

\begin{figure}[t]
\begin{center}
\includegraphics[width=.48\textwidth]{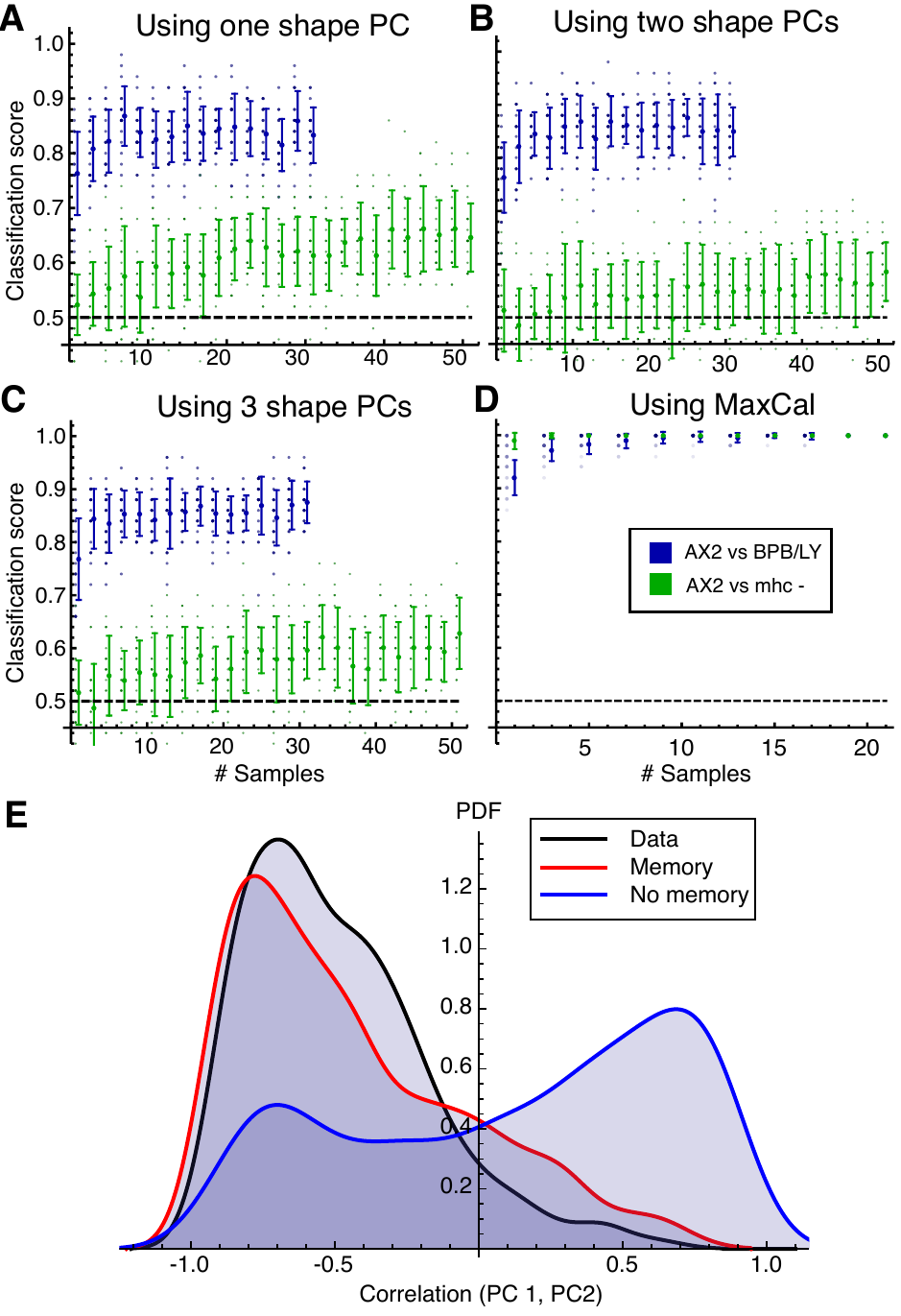}
\caption{{\bf Discriminators trained on MaxCal behavioral descriptors perform better than shape alone.} {\bf(A-C)} Success rates for \emph{k}-nearest neighbors based on the shapes of cells alone. Some number of samples are taken from the same condition and individually classified using the first 1 (A), 2 (B) or 3 (C) shape PCs, with the most commonly chosen classification used to identify the group. Classification is performed on both untreated vs BPB/LY treated (blue) and parental vs \( \emph{mhcA}^{-} \) (green) data. {\bf (D)} A similar \emph{k}-nearest neighbor classification to (A-C), but with distances between neighbors based on the values of the two Lagrange multipliers that best separate the training data.  {\bf (E)} Degree of correlation between PC 1 and PC 2 across a number of simulated cell trajectories. Three lines are shown, corresponding to the data (black), a non-Markovian (red) master equation, in which transition probabilities depend on the last transition made, and a Markovian (blue) master equation, in which they simply depend on the current state. The Markovian model (without memory) lacks the skew toward negative correlations seen in the data. The non-Markovian model (with memory) recovers these.}
\label{fig:shapeIntro}
\end{center}
\end{figure} 

\end{document}